# Direct Observation of Valley-polarized Topological Edge States in Designer Surface Plasmon Crystals

Xiaoxiao Wu[1,2,a], Yan Meng[1,2,a], Jingxuan Tian[3], Yingzhou Huang[1], Hong Xiang[1], Dezhuan Han[1,b], Weijia Wen[1,2,b]

[1]*Department of Applied Physics, Chongqing University, Chongqing 401331, China*

[2]*Department of Physics, The Hong Kong University of Science and Technology, Clear Water Bay, Kowloon, Hong Kong, China*

[3]*Department of Mechanical Engineering, Faculty of Engineering, The University of Hong Kong, Hong Kong, China*

[a] Xiaoxiao Wu and Yan Meng contributed equally to this work.

[b] Correspondence and requests for materials should be addressed to Dezhuan Han (email: dzhan@cqu.edu.cn) or Weijia Wen (email: phwen@ust.hk).

## Abstract

The extensive research of two-dimensional layered materials has revealed that valleys, as energy extrema in momentum space, could offer a new degree of freedom for carrying information. Based on this concept, researchers have predicted valley-Hall topological insulators which could support valley-polarized edge states at non-trivial domain walls. Recently, several kinds of photonic or sonic crystals have been proposed as classical counterparts of valley-Hall topological insulators. However, direct experimental observation of valley-polarized edge states in photonic crystals is still difficult until now. Here, we demonstrate a designer surface plasmon crystal comprising metallic patterns deposited on a dielectric substrate, which can become a valley-Hall photonic topological insulator by exploiting the mirror-symmetry-breaking mechanism. Topological edge states with valley-dependent transport are directly visualized in the microwave regime. The observed edge states are confirmed to be fully valley-polarized through spatial Fourier transforms. Topological protection of the edge states at sharp corners is also experimentally demonstrated.

**Introduction**

Two-dimensional (2D) layered materials, such as graphene and transition metal dichalcogenides (TMDs), exhibit a pair of degenerate but inequivalent energy extrema in the momentum space (or reciprocal space), called valleys[1-4]. As a new potential carrier for information, valleys have opened up a novel research area referred to as valleytronics[5-9] aimed at engineering the degree of freedom. Many works have predicted that if the degeneracy between the two valleys is lifted, 2D materials could exhibit quantum valley-Hall effect, which is manifested by a pair of counter-propagating edge states with opposite valley-polarization at non-trivial domain walls in the absence of inter-valley scattering[10-14]. These materials are sometimes referred to as valley-Hall topological insulators[14,15]. Regarding the classical counterparts of this kind of quantum topological electronic system, several photonic crystals have been proposed as possible valley-Hall photonic topological insulators (PTIs) and numerically investigated[16-19]. Moreover, a sonic crystal comprising anisotropic scatters has been proposed and experimentally demonstrated as a valley-Hall sonic topological insulator (STI)[20].

The topological phase transition in the STI is induced by mirror symmetry breaking in the sonic crystal. However, currently proposed PTIs and STIs usually require covers to confine the waves in the vertical direction to prevent radiative leakage into free space[20-25]. In experiments, the covers will largely hamper probing inside the photonic or sonic crystals and the experimental mapping of the fields inside the structure becomes a difficult challenge. As a result, there are few direct experimental observations of valley-polarized edge states. Until now, valley-dependent transport has been experimentally reported in STIs[20,26], serving as an important but indirect observation for valley-polarized edge states. In order to achieve direct observations, the edge states need to be vertically confined in free space to eliminate covers[27], and the elimination of covers is also beneficial for both manufacturing and applications.

In view of these demands, designer surface plasmons (DSPs), also dubbed spoof surface plasmons[28-39], can become a potential platform for constructing valley-Hall PTIs. DSPs are non-leaky electromagnetic (EM) surface modes similar to the famous surface plasmons at optical and infrared regimes but arise at much lower frequencies in periodic subwavelength metallic structures, and these structures can be called designer surface plasmon (DSP) crystals. Owing to its vertical confinement in free space, DSPs have been successfully employed in realizing time-reversal-invariant PTIs aimed at directly detecting topological phenomena of classical waves in recent works[40-42]. Compared with previous PTIs which utilize magneto-optical effect[21,43], PTIs based on DSP crystals are time-reversal-invariant and do not rely on external magnetic fields. Further, since the realization of a topological bandgap now does not involve coupled-resonator optical waveguides (CROWs) or the coupling between transverse-electric (TE) and transverse-magnetic (TM) modes, complex structures such as ring-resonator networks[44,45] or 3D resonators[22,23,46] in previous time-reversal-invariant PTIs are no longer needed, which can offer more compactness for the PTIs in applications.

In this paper, we experimentally realize a valley-Hall PTI operating in the microwave regime and demonstrate the measured near-field maps of the valley-polarized edge states. The valley-Hall PTI is constructed using a DSP crystal, and the topological phase transition is realized through perturbations which break mirror symmetry and generate non-trivial valley Chern numbers. Valley-polarized edge states and their valley-dependent transport are directly observed in a beam splitter. The topological protection of these edge states at sharp corners is also demonstrated in a Z-shaped waveguide. The valley-Hall PTI should have potential in applications considering its planar geometry and ultrathin thickness and may serve as a prototype for future development of telecommunication devices.

## Results

**Realizing topological phase transition and valley-Hall PTI using DSP crystal.** The starting point for our consideration is a triangular-lattice DSP crystal comprising metallic patterns deposited on a dielectric substrate as illustrated in Fig. 1(a). The substrate is a conventional high-frequency dielectric material, F4B, with a relative permittivity 2.65 and a loss tangent 0.001. The lattice constant is $a = 12$ mm and the thickness of the substrate is $t = 1$ mm. A top-view Wigner-Seitz unit cell of the DSP crystal is shown in Fig. 1(b). The metallic pattern in each unit cell is a thin layer of copper with thickness $t_m = 35$ μm, comprising an inner circle with a radius $r = 1$ mm and six radial fans with a sector angle of 30°. Of the six fans, the three marked by red outlines have the radius $R_1$ and the other three marked by blue outlines have the radius $R_2$. The first Brillouin zone (FBZ) of the triangular lattice is depicted in Fig. 1(c).

We first consider a detailed situation when the unit cell possesses mirror symmetry along ΓK direction (ΓK-mirror symmetry) and assume that $R_1 = R_2 = 5.55$ mm. The band structure of the DSP crystal is numerically calculated (see Methods) and the guided part[27,47] of the first three bands along special directions of the FBZ is shown in Fig. 1(d), while the dispersion of electromagnetic (EM) wave in air (light line) is represented by the blue dashed lines. The 1st band is highlighted by the red solid line and other bands are represented by black solid lines. It can be observed that the 1st and 2nd bands linearly touch each other at the Dirac frequency $\omega_D = 2\pi \times 7.61$ GHz at the K and K' valleys due to the $C_{3v}$ symmetry of the valleys in the reciprocal space[48]. Dirac cones are then formed at each valley by the two bands, with details displayed in the inset which enlarges the area enclosed by the green rectangle.

We then plot the cross-sectional field maps of $E_z$ component of the two degenerate states at the K valley in Fig. 1(e), where the blue arrows represent the time-averaged Poynting vectors (energy flux). As displayed in the field maps, the energy flux of the two states are either left circularly polarized (LCP) or right circularly polarized (RCP), and are invariant under a rotation through 120°. The fields at the K' valley could be obtained by the time-reversal operation, which reverses the direction of the energy flux and hence the circular polarization of the state.[16,49] Since the valley states are in the guide part of the band structure and far away from the light line, the electric field is tightly confined near the metallic surface in the $z$ (vertical) direction (See Supplementary Fig. 1 for electric field distribution in the $z$ direction).

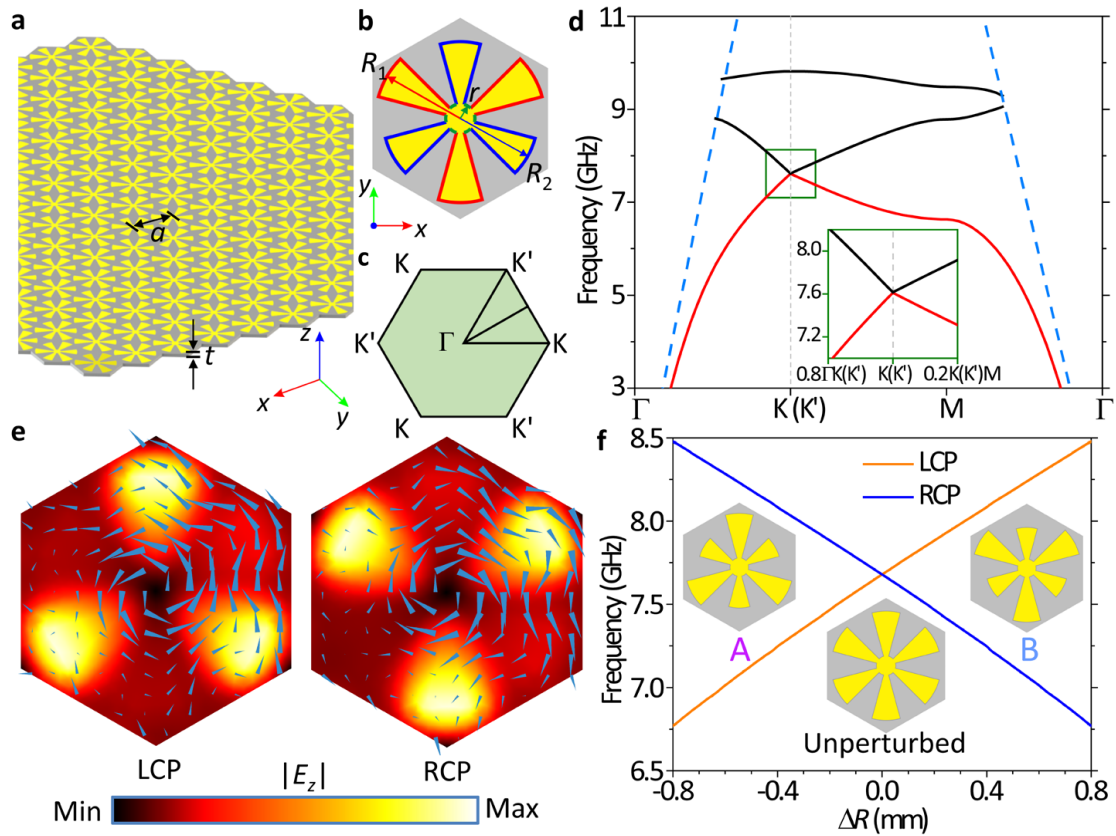

**Figure 1 | Conceptual illustration of the designer surface plasmon (DSP) crystal and frequency split of circularly polarized valley states.** (a) Schematic illustration of the triangular-lattice DSP crystal consisting of metallic patterns (yellow) on a dielectric

F4B substrate (grey). (b) Top-view Wigner-Seitz unit cell of the lattice. The metallic pattern consists of an inner circle with radius $r$ = 1mm and six radial fans with sector angle 30°, three of them marked by red outlines having radius $R_1$, the other three marked by blue outlines having radius $R_2$. (c) First Brillouin Zone (FBZ) of the triangular lattice. (d) Calculated band structure of the DSP crystal along the special directions of the FBZ with $R_1 = R_2 = 5.55$ mm. The blue dashed lines represent the dispersion of EM wave in air (light line) and only the guided part of the band structure is shown. The inset enlarges the area enclosed by the green rectangle around K/K' valley, showing the gapless Dirac cone. (e) Simulated field maps of $E_z$ component on the $xy$ plane 1 mm above the surface, showing the two degenerate states at the K valley with energy flux represented by blue arrows. The circulating energy flux of the states is either left-hand circularly polarized (LCP) or right-hand circularly polarized (RCP). The color indicates the amplitude of $E_z$ component. (f) Calculated eigenfrequencies of the LCP and RCP states at the K valley when the parameter $\Delta R = R_1 - R_2$ varies while the mean radius $\bar{R} = (R_1+R_2)/2 = 5.55$ mm is kept constant. The inversion of frequency order of the LCP and RCP states indicates a topological phase transition when $\Delta R$ crosses zero. The evolution of the unit cell is indicated in the inset, where pattern A corresponds to $\Delta R = -0.5$ mm, the unperturbed pattern corresponds to $\Delta R = 0$, and pattern B corresponds to $\Delta R = 0.5$ mm. The difference between $R_1$ and $R_2$ are exaggerated for a clearer visualization.

We then introduce a perturbation to break the ΓK-mirror symmetry of the unit cell,

which will lift the degeneracy at the valleys and hence open a bandgap[20,49]. This could be accomplished by perturbing the radii $R_1$ and $R_2$ such that they become unequal, and hence a parameter $\Delta R = R_1 - R_2$ is defined to characterize the perturbation. After varying this parameter, the eigenfrequencies of the LCP and RCP states at the K valley are calculated and plotted as a function of $\Delta R$ in Fig. 1(f). For comparison, the mean radius $\bar{R} = (R_1+R_2)/2$ is kept constant ($\bar{R} = 5.55$ mm) when we vary $\Delta R$. The evolution of the unit cell when varying $\Delta R$ is visualized in the inset in Fig. 1(f), pattern A corresponding to $\Delta R = -0.5$ mm, the unperturbed pattern corresponding to $\Delta R = 0$ mm, pattern B corresponding to $\Delta R = 0.5$ mm. As expected, the LCP and RCP states become non-degenerate when $\Delta R$ becomes non-zero and the degeneracy is lifted. However, the $C_3$ symmetry is intact under the perturbation; hence the circular polarizations of the valley states are not mixed[16]. More interestingly, the frequency order of the LCP and RCP states at the K valley is inverted when $\Delta R$ crosses zero, which signals a topological phase transition.

For further analysis, we first study the band structure of pattern A with $\Delta R = -0.5$ mm, that is, $R_1 = 5.3$ mm and $R_2 = 5.8$ mm. Because of the time-reversal symmetry as no magnetic materials are introduced, the band structures of pattern A and pattern B are exactly the same, and the guided part of their band structure is shown in Fig. 2(a). As can be seen, the full bandgap indicated by the yellow shaded region confirms that the breaking of ΓK-mirror symmetry indeed lifts the degeneracy between the 1st band and the 2nd band at the valleys. Further, it is noted that though pattern A and pattern B share the same band structure, the circular polarizations of their energy flux at the K and K' valleys are exactly opposite (See Supplementary Fig. 2) as dictated by the time-reversal symmetry.

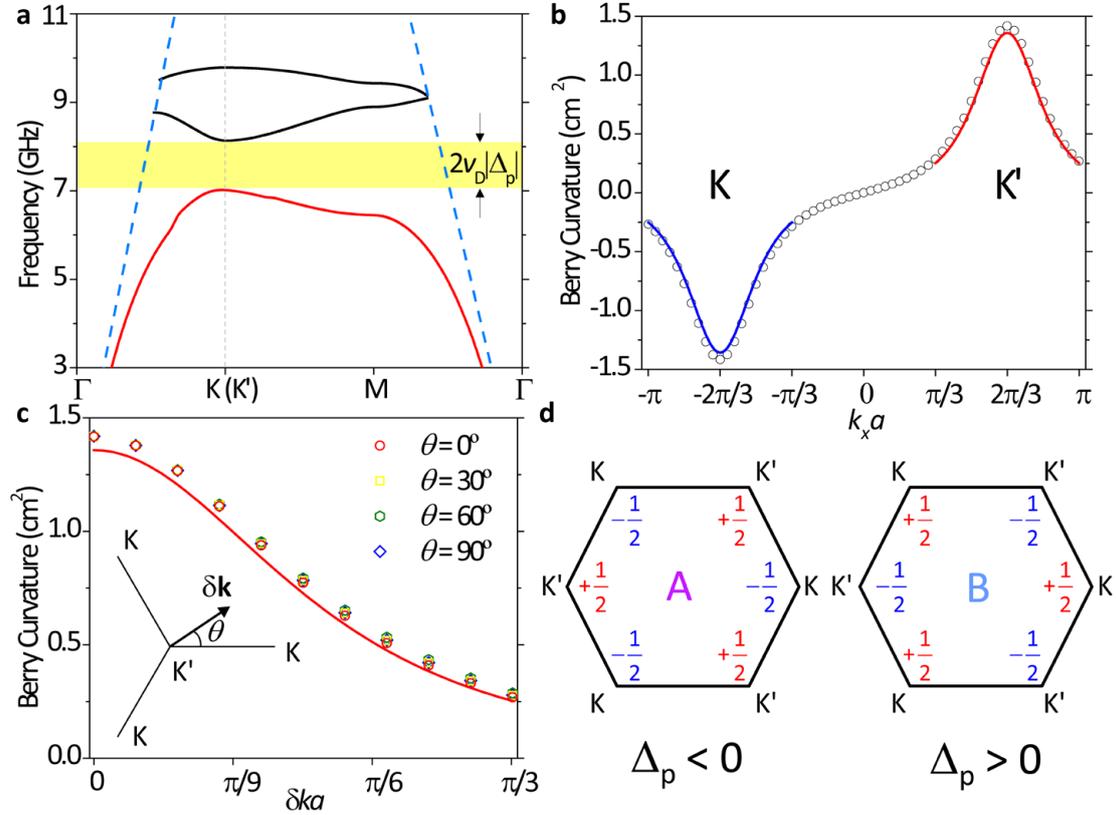

**Figure 2 | Band structure of the DSP crystal with breaking of ΓK-mirror symmetry and topological property of its 1st band.** (a) Calculated band structure of the DSP crystal along the special directions of the FBZ when Δ$R$ = ±0.5 mm (pattern A or B). Blue dashed lines represent the light line and only the guided part of the band structure is shown. The light yellow shaded region represents the full bandgap. (b) Berry curvature of the 1st band for pattern A along the line $k_y = 2\pi/(\sqrt{3}a)$ in the reciprocal space from numerical calculations (open circles) and effective Hamiltonian model near the K (blue solid line) or K' (red solid line) valley. (c) Berry curvature for pattern A around a K' valley as a function of δ$k$ for different polar angles $\theta$ (open symbols). The red solid line represents the isotropic result from the effective

Hamiltonian model. The inset shows the definition of δk and θ in the reciprocal space. (d) Valley Chern numbers summarized for pattern A and pattern B.

**Theoretical modeling of topological phase transition and numerical verifications.** To model the topological phase transition from pattern A to pattern B, which could not be fully investigated from the band structure, we construct an effective Hamiltonian using **k·p** method in the vicinity of K/K' valley[50,51] (See Supplementary Note 1 for the first-principal derivation)

$$H_{K/K'}(\delta \mathbf{k}) = \pm v_D \delta k_x \sigma_x + v_D \delta k_y \sigma_y + v_D \Delta_p \sigma_z, \tag{1}$$

in which $v_D$ is the group velocity, $\delta \mathbf{k} = \mathbf{k} - \mathbf{k}_{K/K'}$ is the displacement of the wave vector **k** to the K/K' valley represented by $\mathbf{k}_{K/K'}$ in the reciprocal space, $\sigma_i$ ($i = x, y, z$) are the Pauli matrices acting on the orbital degree of freedom, and $\Delta_p$ is proportional to the bandwidth of the bandgap ($2v_D|\Delta_p|$) between the 1st and 2nd bands (See Methods). Using the eigenvector from this effective Hamiltonian, the local Berry curvature (z component) of the 1st band at the K/K' valley can be analytically calculated (See Supplementary Note 2)

$$\Omega_{z,K/K'}(\delta \mathbf{k}) = \pm \frac{\Delta_p}{2(\delta k^2 + \Delta_p^2)^{3/2}}, \tag{2}$$

in which $\delta k = |\delta \mathbf{k}|$ is the norm of the displacement in the reciprocal space, and the result is expectedly localized around the valleys ($\delta k = 0$). In order to consolidate the above analytical results, we numerically calculate the Berry curvature along the line $k_y = 2\pi/(\sqrt{3}a)$ in the reciprocal space which includes a pair of K and K' valleys (see

Methods). The numerical result is shown as open circles in Fig. 2(b). We also analytically calculate the Berry curvature according to Eq. (2) with fitted parameters ($v_D$ = 5.42×10$^7$ m·s$^{-1}$ and $\Delta_p$ = −60.68 m$^{-1}$, see Methods), and this result, which is essentially derived from the effective Hamiltonian model near K/K' valley, is separately plotted as the blue and red solid lines in Fig. 2(b). Quantitative agreement can be observed between the numerical calculation and effective Hamiltonian model. However, the anisotropy of the local Berry curvature is not embodied in Eq. (2) because Eq. (1) is a lowest-order effective Hamiltonian which only includes the linear isotropic terms, and higher order anisotropic terms, such as warping terms, are neglected.

To evaluate the anisotropy, we numerically calculate the Berry curvature around a K' valley along lines in the reciprocal space with different polar angle $\theta$, and the corresponding results are plotted as a function of $\delta k$ in Fig. 2(c), represented by differently colored open symbols. For comparison, we also plot the isotropic result from the effective Hamiltonian model as the red solid line. From the results, we could conclude that the anisotropy is indeed negligible and the Berry curvature calculated from the effective Hamiltonian is satisfactorily accurate. Therefore, we can safely calculate the valley Chern numbers of the 1st band by integrating the local Berry curvature derived from the effectively Hamiltonian

$$C^{K/K'} = \frac{1}{2\pi}\int_{HBZ} \Omega_{K/K'}(\delta \mathbf{k})dS = \pm\frac{1}{2}\text{sgn}(\Delta_p), \quad (3)$$

in which the integral is carried out in half of a Brillouin zone (HBZ) surrounding the K/K' valley. Eq. (3) shows that for each valley the valley Chern number is ±1/2 with the sign solely determined by the sign of $\Delta_p$. This dependence stems from the fact that the sign of $\Delta_p$ solely determines the frequency order of the LCP and RCP states in the band structure, and their frequency order determines the topological phase of the DSP crystal. The calculation using Eq. (3) is straightforward since $\Delta_p$ is negative for pattern A and hence positive for pattern B. The calculated valley Chern numbers for pattern A and pattern B are summarized in Fig. 2(d). These results indicate that when $\Delta R$ is nonzero, the proposed DSP crystal becomes a valley-Hall PTI because of the full bandgap and nontrivial valley Chern numbers[16,19].

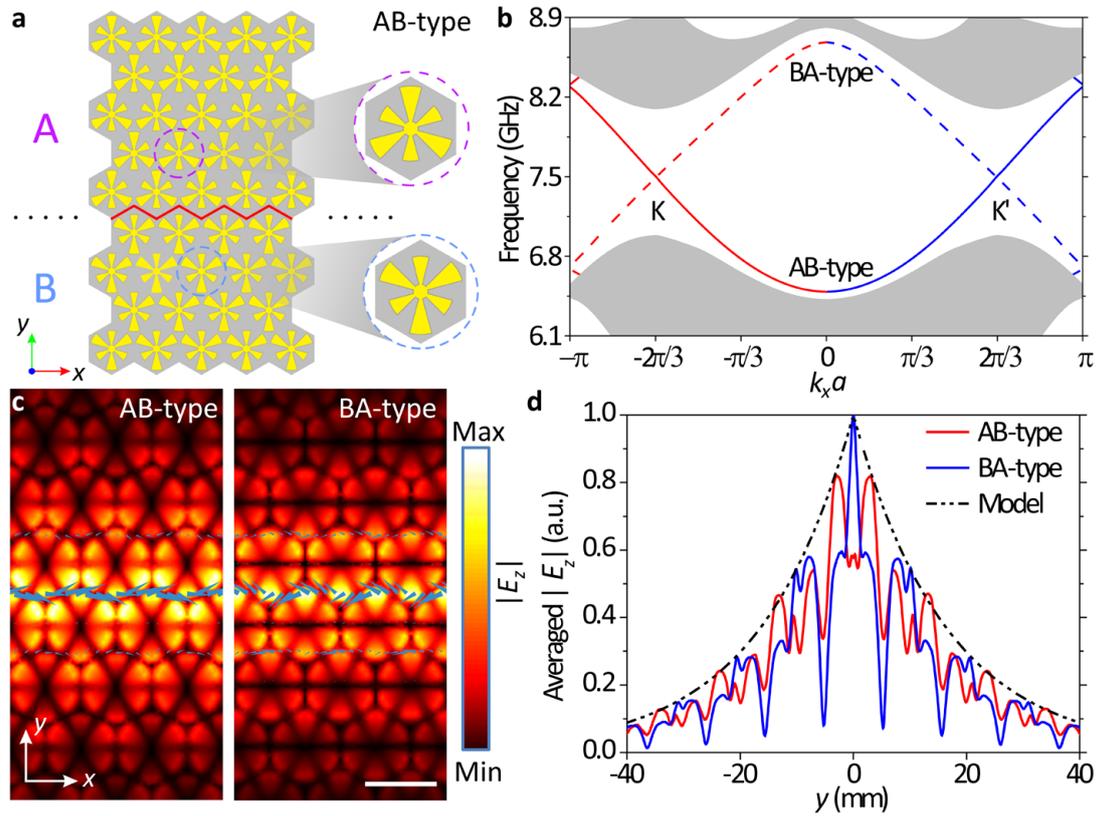

**Figure 3 | Edge states at domain walls between two valley-Hall PTIs.** (a) Sketch of an AB-type domain wall, in which pattern A is replicated at the upper domain and pattern B is replicated at the lower domain, respectively. The domain wall is marked by the red solid lines. (b) Calculated band structure of an AB-type (solid line) domain

wall and a BA-type (dashed line) domain wall. The shaded regions represent the projection of the bulk bands. The red lines indicate positive group velocity (forward) and the blue lines indicate negative group velocity (backward). (c) Simulated field maps of $E_z$ component on the *xy* plane 1 mm above the surface, showing the edge states of AB-type (left panel) domain wall and BA-type (right panel) domain wall at the K' valley, respectively. The color indicates the amplitude of $E_z$ component. The blue arrows represent the energy flux. The scale bar is 12 mm. (d) The averaged $|E_z|$ (averaged along the *x* direction in a unit cell) of the edge states shown in (c). The black dash-dotted line represents the corresponding result calculated from the effective Hamiltonian model.

**Bulk-boundary correspondence and emergence of valley-polarized topological edge states.** The above conclusion could be reinforced by the famous bulk-boundary correspondence, which states that the valley topological property of a bandgap is determined by the sum of the valley Chern numbers of the bands below it[52]. Therefore, non-trivial edge states should emerge in the 1st bandgap at a domain wall between pattern A and pattern B, because the difference of valley Chern numbers of their 1st band gives $|\Delta C^{K/K'}| = |C_A^{K/K'} - C_B^{K/K'}| = 1$ [16,19]. We hence combine pattern A and pattern B to construct a domain wall as shown in Fig. 3(a), which is referred to as AB-type because pattern A is replicated in the upper domain and pattern B is replicated in the lower domain. Alternatively, we can also replicate pattern A in the lower domain and pattern B in the upper domain, which forms a domain wall referred to as BA-type. It should be emphasized that an AB-type domain wall cannot be converted to a BA-type domain wall through SO(3) operations, and vice versa. For convenience, we assume that the domain wall is periodic along the *x* direction and centered along *y* = 0.

The band structures of the two domain walls are then simulated using a supercell consisting of total 25 unit cells in the *y* direction (see Methods) and the simulated band structure is shown in Fig. 3(b), in which the shaded regions represent the projections of bulk bands. As expected, gapped edge states are observed in the 1st bandgap for each domain wall, and they do not connect the 1st and 2nd bands because time-reversal symmetry is preserved here[19]. The edge states of the AB-type and BA-type domain walls are represented by the solid and dashed lines, respectively, and the edge states of the two domain walls intersect each other at the K/K' valley at the frequency 7.50 GHz, in the middle of the 1st bandgap (See Supplementary Note 3). We use red and blue colors to denote the part of the edge states near K and K' valleys, respectively.

It is seen that the propagating directions of an edge state for K and K' valleys are exactly opposite, no matter at an AB-type or BA-type domain wall. This is a manifestation of valley-chirality of the valley-Hall effect, in other words, the valley-polarized edge states are locked to one propagating direction in the absence of inter-valley scattering[14,20]. For example, at the K valley, the edge state of the AB-type domain wall is backward, while at the K' wall it is forward. This phenomenon could be predicted by the valley Chern number, as $\Delta C_{AB}^{K} = C_{A}^{K} - C_{B}^{K} = -1$ and $\Delta C_{AB}^{K'} = C_A^{K'} - C_B^{K'} = 1$. The simulated field maps of $E_z$ component of edge states at the K' valley are plotted in Fig. 3(c), and both map show that the field is concentrated along the domain wall and decays into the bulk, which is a characteristic of edge states. Further, the blue arrows represent the energy flux and their directions verify that at the K' valley, the edge state of AB-type is forward while that of BA-type is backward, consistent with the band structure in Fig. 3(b).

We also use three phase-matched dipoles[16] to excite the edge states at domain walls, and the observed one-way propagation in simulations and experiments confirms the valley chirality of the edge states (See Supplementary Note 5 and Supplementary Figs. 5 and 6). Then in order to analyze the decay character of the edge states in the bulk, $|E_z|$ is averaged along the $x$ direction in a unit cell and plotted as a function of $y$ in Fig. 3(d). It can be observed that the averaged $|E_z|$, which represents spatial amplitude profiles of the edge states, symmetrically decay into the bulk with respect to $y = 0$.

The analytic form of the edge states can also be directly deduced from the effective Hamiltonian model, which proves the bulk-boundary correspondence in our structure, and the decay of the edge state is predicted to be $|E_z| \propto e^{-\Delta_p |y|}$ (See Supplementary Notes 3). As can be observed in Fig. 3(d), the prediction from the effective Hamiltonian model plotted as the black dashed line agrees well with the envelopes of the averaged $|E_z|$ from simulations. Therefore, the symmetric decay of the edge states should be attributed to the same $\Delta_p$, or the same decay length in the bulk, shared by pattern A and pattern B lying on the two sides of the domain walls.

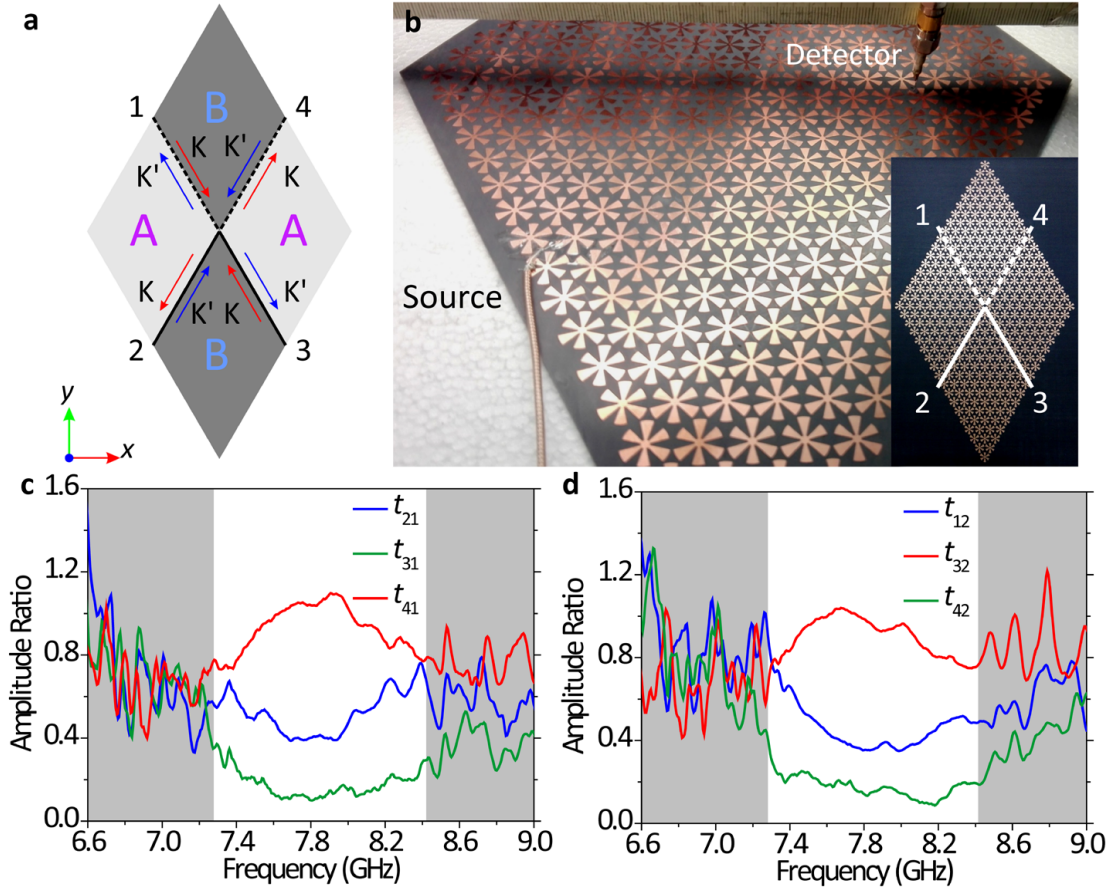

**Figure 4 | Topological transport of valley-polarized edge states.** (a) Schematic diagram of the beam splitter used for selective excitation of valley-polarized edge states. The beam splitter comprises two domains of patterns A and two domains of patterns B. The black dashed lines and solid lines represent AB-type domain walls and BA-type domain walls, respectively. The red and blue arrows represent edges states near K and K' valleys, respectively. (b) Photograph of the experiment setup. Two monopole antennas are used in experiments (see Methods). DSPs are excited by one of the antennas and $E_z$ component of the transmitted wave is measured by the other one. Inset shows a top-view photograph of the fabricated beam splitter. (c), (d) Measured $E_z$ amplitude ratio $t_{ij}$ (outgoing terminal $i$, incident terminal $j$) when the source is placed at terminal 1 (c) or terminal 2 (d). The unshaded region corresponds

to the bandgap obtained from experimental measurement.

**Experimental measurement of valley-dependent transport in a beam splitter**. In order to selectively excite valley-polarized edge states, a rhombus-shaped beam splitter is designed and its diagram is shown in Fig. 4(a). The beam splitter consists of four domains. The upper and lower domains are filled with patterns A, while the left and right domains are filled with patterns B, with each domain containing 9×9 unit cells. Between the domains, two AB-type domain walls and two BA-type domain walls are formed and they are denoted by the dashed lines and solid lines, respectively. At the outer ends of the domain walls, there exist four terminals labeled as 1-4. The valley-polarized edge states supported at the four domain walls are indicated in Fig. 4(a), in which red or blue arrows imply K or K' valley-polarized edge states.

A real sample of the beam splitter is fabricated by etching a 35-μm thick copper layer on a 1-mm thick F4B substrate using standard printed circuit board (PCB) etching techniques. The experiment setup is shown in Fig. 4(b), in which the source and detector are connected to port 1 and port 2 of a vector network analyzer and the scattering parameter $S_{21}$ proportional to the $E_z$ component is measured (see Methods). The inset of Fig. 4(b) shows a top-view photograph of the fabricated sample and indicates the domain walls and terminals of the sample. The measured $S_{21}$ data are regarded as $E_z$ and the amplitude ratios of $S_{21}$ between the outgoing terminals and the incident terminal are plotted in Fig. 4(c) and 4(d), in which the source is attached at terminal 1 and terminal 2, respectively. The measured $E_z$ amplitude ratios are labeled as $t_{ij}$ by the outgoing terminal $i$ and incident terminal $j$. The unshaded region corresponds to the measured bandgap of the fabricated DSP crystal samples (See Supplementary Fig. 3 for measurement of the bandgap). Compared with the simulated bandgap (7.05-8.12 GHz), the measured bandgap (7.28-8.42 GHz) is shifted about 3.7% towards higher frequency, and this small deviation could be attributed to fabrication errors, such as the errors in thickness and relative permittivity of the commercial F4B substrates. Nevertheless, these fabrication errors do not introduce any essential difference in the physics of valley-dependent transport in our investigated DSP crystal, and only slightly shifts the frequency of bandgap edges (See Supplementary Fig. 4 for detailed results).

From the measured amplitude ratios, it can be observed that the transports of edge states involving opposite valleys ($t_{31}$ in Fig. 4(c) and $t_{42}$ in Fig. 4(d)) are heavily suppressed compared with those involving the same valley in the bandgap. In the bandgap, the sum of the measured $E_z$ amplitude ratio involving the same valley is at least 14 dB larger than those involving opposite valleys, whenever the source is attached at terminal 1 or terminal 2. The measurements hence indicate that the excited edge states of the DSP crystal adhere to a valley-dependent topological transport, a feature of valley-polarized edge states[20]. The total thickness (1.035 mm) of the fabricated DSP crystal is smaller than 1/38 of the wavelength (40 mm) at 7.50 GHz, and this deep-subwavelength thickness could be further decreased if we use a dielectric substrate with higher permittivity.

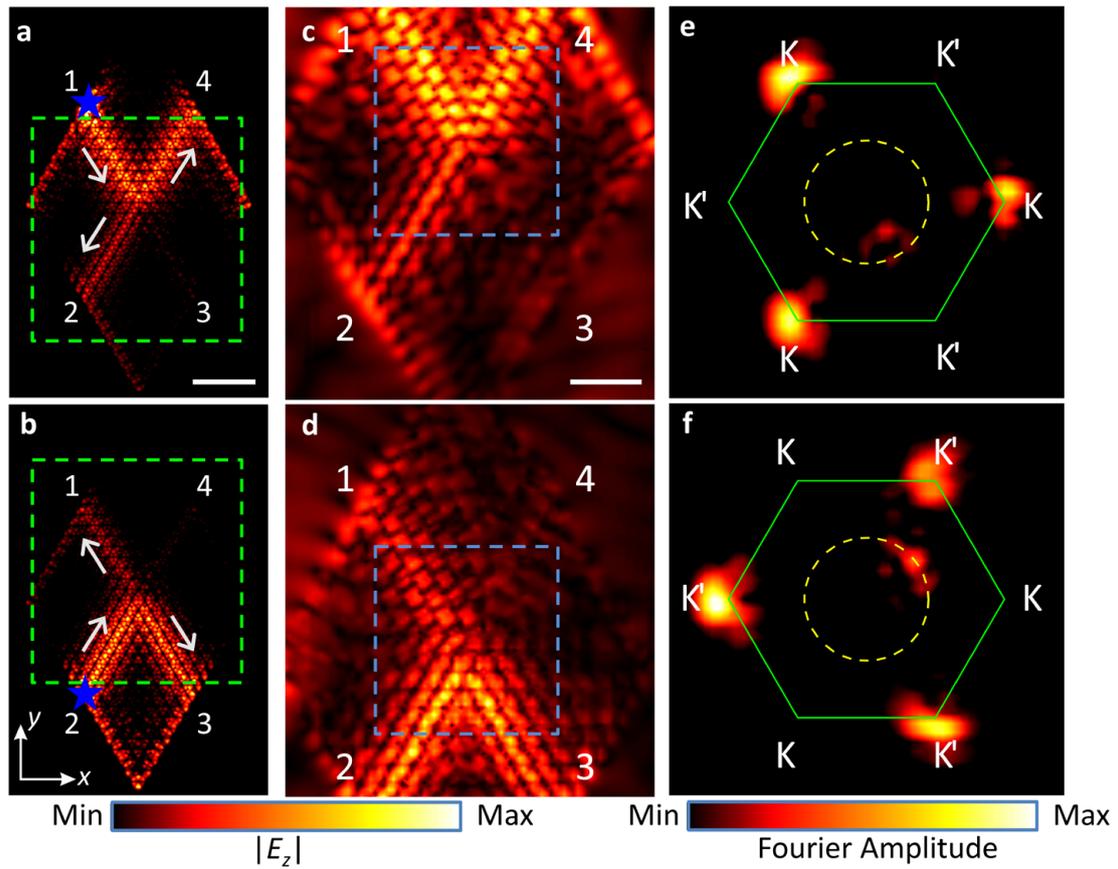

**Figure 5 | Near-field maps of valley-polarized edge states and corresponding spatial Fourier spectra.** (a), (b) Simulated field maps of $E_z$ component at 7.50 GHz on the *xy* plane 1 mm above the metallic surface. The source denoted by the blue star is placed at terminal 1 in (a) and terminal 2 in (b). Grey arrows denote the direction of the valley-polarized edge states. Green dashed rectangles denote the regions of the area 205×220 mm$^2$ scanned in experiments. The scale bar is 60 mm. (c), (d) Measured field maps of $E_z$ component at 7.50 GHz in the region of the sample denoted by green dashed rectangles in (a) and (b), respectively. The blue dashed rectangles denote the interior regions where we perform Fourier transforms. The scale bar is 40 mm. (e), (f) Spatial Fourier spectra of the scanned $E_z$ component. The green solid hexagons represent the FBZs and the yellow dashed circles represent the 7.50-GHz

isofrequency contours of the light cone. The color indicates the amplitude of $E_z$ component in field maps and the amplitude of spatial Fourier transforms in spatial Fourier spectra, respectively.

**Direct mappings of topological edge states and analysis of their valley-polarization.** Next, we scan the $E_z$ component on the *xy* plane above the sample with the same experiment setup for the purpose of directly visualizing the edge states and analyzing their valley polarizations. For reference, simulated field maps of $E_z$ component at 7.50 GHz are plotted in Figs. 5(a) and 5(b), in which the blue star denotes the source and the grey arrows denote the directions of the propagating edge states. The source is placed at terminal 1 and terminal 2, respectively, in Figs. 5(a) and 5(b). We then experimentally scanned the $E_z$ component in the regions denoted by green dashed rectangles in Figs. 5(a) and 5(b). The two regions are both of the area 205×220 mm$^2$. The scanned field maps at 7.50 GHz are plotted in Figs. 5(c) and 5(d), respectively (See Supplementary Fig. 7 for scanned field maps at other frequencies).

The field maps from simulations and experiments show similar distributions of the electric fields and clearly demonstrate the suppression of the inter-valley edge states, which has been indicated in Figs. 4(c) and 4(d). Spatial Fourier transforms[49] are then performed on the interior regions (denoted by the blue dashed rectangles in Figs. 5(c) and 5(d)) of the scanned complex $E_z$ component at 7.50 GHz, which excludes the interfaces between DSP crystals and air. The amplitudes of the spatial

Fourier transforms, or so-called spatial Fourier spectra, are then plotted in Figs. 5(e) and 5(f), respectively, where the green solid hexagons denote the FBZs and the yellow dashed circles denote the 7.50-GHz isofrequency contours of the light cone.

From the spatial Fourier spectra, it is observed that wherever the source is attached at terminal 1 or terminal 2, only three equivalent corners of the FBZ, corresponding to one valley, are bright in each situation. In contrast, the three dark corners of the FBZ clearly imply that the opposite valley which they corresponding to in each situation is heavily suppressed. The small faint regions on the light cone (yellow dashed circles) further imply that the non-zero $t_{31}$ or $t_{42}$ in Figs. 4(c) and 4(d) are largely owing to the existence of the uncoupled EM wave propagating in air, rather than inter-valley scatterings of the edge states.

We have scanned the background field when there are no metallic patterns on the F4B substrate and the corresponding spatial Fourier spectra also show similar faint regions on the light cone (See Supplementary Note 6 and Supplementary Fig. 8 for detailed experiment setup and results). The small faint regions inside the light cone could be attributed to the evanescent states created by the scatterings of the edge states at the center of the beam splitter[53]. The spatial Fourier spectra hence directly confirm that the excited propagating edge states are fully valley-polarized and their valley-polarizations totally agree with the diagram in Fig. 4(a). In other words, the spatial Fourier spectra experimentally confirm the that the observed edge states could propagate with negligible inter-valley scattering along zigzag domain walls, which has been theoretically predicted and numerically confirmed for EM waves in previous works[16]. This feature indicates that we can exploit the valley-Hall PTIs to design planar topological waveguides since the suppression of inter-valley scattering will suppress scattering losses at sharp corners in conventional waveguides owing to the valley-chirality of the edge states. Compared with previous counterparts[21-23], the topological waveguide does not require external magnetic fields or three-dimensional structures for realizing this topological protection, thus benefiting both manufacturing and applications. Moreover, the planar geometry and ultrathin thickness of the topological waveguide will also facilitate its integration with existing electronic systems.

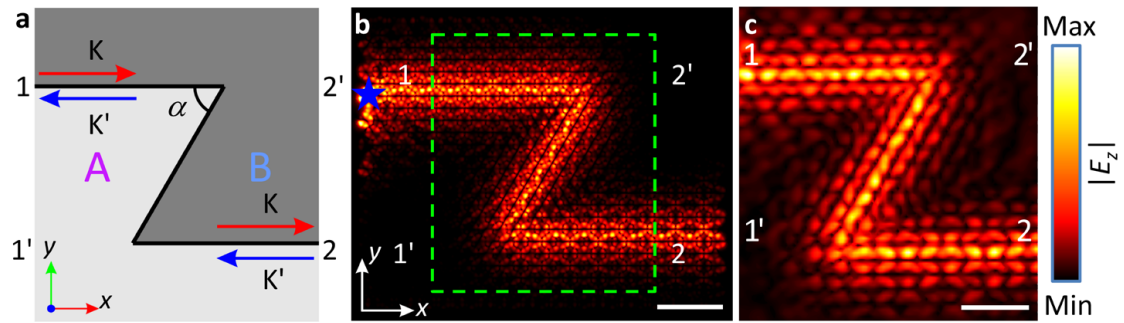

**Figure 6 | Topological protection of the edge state in a Z-shaped waveguide.** (a) Schematic diagram of a Z-shaped topological waveguide formed between patterns A and patterns B. The corner angle $\alpha$ = 60°. (b) Simulated field map of $E_z$ component at 7.50 GHz. The green dashed rectangle denotes the region of the area 208×220 mm$^2$ scanned in experiments. The color indicates the amplitude of $E_z$ component. The scale bar is 60 mm. (c) Measured field map of $E_z$ component at 7.50 GHz. The color indicates the amplitude of $E_z$ component. The scale bar is 40 mm.

**Demonstration of topological protection in a Z-shaped waveguide**. For proof of principle, we designed and fabricated a Z-shaped waveguide featuring two sharp corners using patterns A and patterns B. The schematic diagram of the waveguide is shown in Fig. 6(a), in which the corner angle $α = 60°$. The waveguide is essentially a curved BA-type domain wall and the two ends of the interface are denoted as terminal 1 and terminal 2. In experiments, the source is attached at terminal 1 and the $E_z$ amplitude is measured around terminal 1 and terminal 2. We also measured the $E_z$ amplitude near two reference terminals 1' and 2', which are in fact in the bulk of patterns A or patterns B. The measured amplitude ratios show a good contrast between $t_{21}$ and $t_{1'1}$ or $t_{2'1}$ in the bandgap (See Supplementary Fig. 11(a)), which implies a good confinement of the edge state.

We then scanned the $E_z$ component above the waveguide with the same experiment setup used for scanning the beam splitter. For comparison, the simulated field map of $E_z$ component at 7.50 GHz is shown in Fig. 6(b), in which the green dashed rectangle denotes the region of the area 208×220 mm$^2$ scanned in experiments. The scanned field map at 7.50 GHz is shown in Fig. 6(c), which agrees well with the simulated field map and shows that there are no observable scattering losses at the bends (See Supplementary Fig. 9 for scanned field maps at other frequencies).

We have also measured the transport of the edge state in a straight topological waveguide also comprised of a BA-type domain wall and it is observed that the transport behaviors of the two waveguides are generally similar in the bandgap, but show remarkable differences in the bulk bands (See Supplementary Fig. 11(b)). We have also numerically investigated other geometric configurations in which the corner angle $\alpha$ of the topological waveguides varies, and no notable scattering losses are observed in all cases (See Supplementary Fig. 10 for simulated field maps.), even when the corner becomes incompatible with the triangular lattice ($\alpha$ = 30°, 90°, 150°)[20]. The overall results hence confirm that the corners in the topological waveguides only have weak influences on transport of the edge states, compared with conventional non-topological waveguides[23].

According to previous theoretical studies, a random arrangement of the unit cells of valley-Hall PTIs only results in a higher-order perturbation $O(\Delta_p^2)$ between the two circularly polarized states in K/K' valley, which is usually much smaller than the first-order perturbation $v_D\Delta_p$ caused by breaking ΓK-mirror symmetry[16]. This feature of valley-Hall PTIs leads to the suppression of scattering losses in our topological waveguides, where the unit cells (patterns A and patterns B) are arranged to generate a curved domain wall. Since the underlying mechanism is independent of specific waves, this suppression of scattering losses are also observed for acoustic waves in the topological waveguides made from a valley-Hall STI where the topological bandgap is similarly introduced by breaking the ΓK-mirror symmetry of the unit cell[20].

## Discussion

In summary, we have designed an ultrathin valley-Hall PTI based on mirror-symmetry-breaking mechanism using a DSP crystal. Selective excitation and direct observation of valley-polarized edge states is experimentally demonstrated and confirmed. Our research suggests that DSP crystals which support non-leaky DSPs could become an ideal platform for exploring various phenomena involving topological properties, which are originally proposed in electronic systems. The intrinsic non-leaky feature of DSPs in an open environment facilitates the scanning of near fields, which is difficult for many classical systems, and hence paves the way for more direct experimental studies on topological phenomena of classical waves. Moreover, the proposed valley-Hall PTI could also be useful in telecommunications because of its ease in large-scale fabrication using standard techniques, its planar geometry and ultrathin thickness without any covers, and its topological edge states which can travel through sharp corners with negligible scattering, a critical concern in conventional waveguides. The one-way propagation of valley-polarized edge states may also be exploited to design an on-demand flow switch of EM wave. Finally, we note that though proposed in the microwave regime, the design principle of the valley-Hall PTI should be valid in other frequency regimes, such as the terahertz and near-infrared regimes, and even the optical regime because the basic mirror-symmetry-breaking mechanism does not rely on particular material properties.

## Acknowledgements

The authors would like to thank C.T. Chan and Meng Xiao for fruitful discussions. The work is supported by an Areas of Excellence Scheme grant (AOE/P-02/12) from Research Grant Council (RGC) of Hong Kong and two grants from National Natural Science Foundation of China (NSFC) (No. 11574037, No. 91630205).

## Author Contributions

X. W. and W. W. conceived the original idea. X. W., D. H., and Y. M. performed the simulations and derived the theory. Y. H. supported the fabrication process of the sample. X. W. and Y. M. carried out the experiments. H. X. helped in the experiment setup. X. W., D. H., Y. M. and J. T. analyzed the data, prepared the figures and wrote the manuscript. W. W. and D. H. supervised the project. All authors contributed to scientific discussions of the manuscript.

## Methods

**Effective Hamiltonian Model**

In the effective Hamiltonian, $H_{K/K'}(\delta \mathbf{k}) = \pm v_D \delta k_x \sigma_x + v_D \delta k_y \sigma_y + v_D \Delta_p \sigma_z$, the first two terms describe the gapless Dirac cones at the K/K' valley shown in Fig. 1(b), while the third term comes from the geometric perturbation that breaks the ΓK-mirror symmetry. The third term is also called the mass term[20], which lifts the degeneracy at valleys and opens a full band gap with bandwidth $2v_D|\Delta_p|$ shown in Fig. 2(a). Here $v_D$ is the group velocity, $\delta \mathbf{k} = \mathbf{k} - \mathbf{k}_{K/K'}$ is the displacement of the wave vector $\mathbf{k}$ to the K/K' valley ($\mathbf{k}_{K/K'}$), and $\sigma_i$ ($i = x, y, z$) are Pauli matrices acting on the orbital degree of freedom comprising the 1st and 2nd bands at the K/K' valley. Since the circular polarizations of the 1st and 2nd bands at the K/K' valley are exactly opposite as aforementioned, the two circularly polarized states are represented as $A_R = [1; 0]$, $A_L = [0; 1]$ at the K valley and $A_R = [0; 1]$, $A_L = [1; 0]$ at the K' valley[16], respectively. By fitting the dispersions around the valleys calculated from the effective Hamiltonian to the numerical band structures shown in Figs. 1(b) and 2(a), it is found that for pattern A, the parameters $v_D = 5.42 \times 10^7$ m·s$^{-1}$ and $\Delta_p = -60.68$ m$^{-1}$, both quantitatively agreeing with the values obtained from the first-principal derivation and the deviation is smaller than 1% (See Supplementary Note 1). Likewise, for pattern B, $v_D = 5.42 \times 10^7$ m·s$^{-1}$ and $\Delta_p = 60.68$ m$^{-1}$.

**Simulations**

Throughout this paper, all full-wave simulations (or numerical calculations) are performed using commercial finite element method software COMSOL Multiphysics, in which the module Electromagnetic Waves, Frequency Domain (emw) is used. When calculating the bulk (edge) band structure, Floquet periodic boundary conditions are imposed on periodic surfaces of the unit cell (supercell). Second order scattering boundary conditions for plane waves are used to terminate the simulation domain. In all simulations, the maximum scale of the mesh is smaller than 1/20 of the wavelength (40 mm) of 7.50-GHz EM wave in air. All *xy*-plane field maps are recorded at the plane 1 mm above the metallic surface.

**Numerical Calculations of Berry Curvature**

When numerically calculating Berry curvature of a point in the reciprocal space, we first integrate Berry connection along an infinitesimal square contour around the point. Then, according to Stokes' theorem, the line integral is equal to the surface integral of Berry curvature over the infinitesimal square which includes the point.[54] Hence the line integral divided by the area of the infinitesimal square is exactly the Berry curvature of the point. In real numerical calculations, the side length of the square contour is chosen to be $\delta k = 1.0$ m$^{-1}$, and the path integrals are discretized into summations. Further details are given in Supplementary Note 4 and Supplementary Fig. 12.

**Experiment Setup**

In experiments, two monopole antennas are used to act as the source and the detector, and the sample is supported by a piece of foam which is mounted on a stepper motor-driven stage. A vector network analyzer (VNA, Agilent N5232A) is employed throughout the measurements and the source connected to port 1 of the VNA is closely attached to the surface of the sample to excite DSPs, while the detector connected to port 2 of the VNA are placed ~1 mm above the metallic surface of the sample to measure the $E_z$ component of the electric field. In measurements, the scattering parameter $S_{21}$ is recorded, which is proportional to the $E_z$ component of the electric field. When scanning the $E_z$ component above the sample, the resolution in the $xy$ plane is 1×1 mm$^2$.

**Data Availability**

The data which support the figures and other findings within this paper are available from the corresponding authors upon request.

## Additional Information

Supplementary information is available in the online version of the paper.

# Competing Financial Interests

The authors declare no competing financial interests.